\documentclass [aip, apl, amsmath, amssymb, twocolumn, letterpaper]{revtex4}

\usepackage{amsmath}
\usepackage{graphicx}

\begin{document}

\title{Conductance asymmetry in point-contacts on epitaxial thin films of Ba(Fe$_{0.92}$Co$_{0.08}$)$_2$As$_2$ }

\author{M. Mehta,$^1$ G. Sheet,$^1$ D. A. Dikin,$^1$ S. Lee,$^2$ C.W. Bark,$^2$ J. Jiang,$^3$\\
J. D. Weiss,$^3$ E. E. Hellstrom,$^3$ M. S. Rzchowski,$^4$ C.B. Eom$^2$ and V. Chandrasekhar$^1$ }

\affiliation{$^1$Department of Physics and Astronomy, Northwestern University, Evanston, IL 60208, USA,\\ 
$^2$Department of Materials Science and Engineering University of Wisconsin-Madison, Madison, WI 53706, USA,\\ 
$^3$Applied Superconductivity Center, National High Magnetic Field Laboratory, Florida State University, Tallahassee, FL 32310, USA \\
$^4$Physics Department, University of Wisconsin-Madison, Madison, WI 53706}

\begin{abstract}
One of the most common observations in point-contact spectra on the recently discovered ferropnictide superconductors is a large conductance
 asymmetry with respect to voltage across the point-contact. In this paper we show that the antisymmetric part of the point-contact spectrum between
 a silver tip and an epitaxial thin film of Ba(Fe$_{0.92}$Co$_{0.08}$)$_2$As$_2$ shows certain unique features that have an interesting evolution
 with increasing temperature  up to a temperature far above the critical temperature $T_c$. We associate this observation with the existence of a
 gap above $T_c$ that might originate from strong fluctuations of the phase of the superconducting order parameter.

\end{abstract}

\maketitle

Determining the order parameter symmetry is perhaps the most fundamental problem in understanding the nature of exotic superconductors. 
Point-contact spectroscopy (PCS), owing to its ability to provide both energy and momentum-resolved spectroscopic information, has proved to be an 
extremely powerful tool in determining the order parameter symmetry in unconventional superconductors. The list includes $d$-wave
superconductors like the cuprates \cite{Deutscher_RMP}, multiband superconductors like MgB$_2$ \cite{szabo_Mgb2} and the borocarbides
\cite{Goutam_BC}. Apart from determining the order parameter symmetry, PCS has also been applied to understanding 
important Fermi-surface properties in the heavy-Fermion superconductors \cite{Greene_HF}. Early PCS experiments on the recently
discovered ferropnictide superconductors indicated the existence of a single BCS (Bardeen-Cooper-Schrieffer)-like superconducting gap \cite{Chen}.
More recent measurements reported the existence of features associated with multiple superconducting gaps possibly originating from the different 
regions of the disjoint Fermi surface of this class of materials \cite{Gonnelli, Samuely}. However, one of the common features in the majority of 
the point-contact spectra reported on the ferropnictides is a large asymmetry in the conductance with respect to voltage across the 
point-contact \cite{Chen, Gonnelli}. In the past, a similar asymmetry was observed in point-contacts involving different complex superconducting systems.
Such conductance asymmetry has been attributed to the Fano resonance in the case of heavy fermion systems \cite{Greene_HF} and unusual Fermi
surface characteristics in the case of the ferropnictides \cite{Gonnelli2}. However, a systematic study of the asymmetry in the ferropnictides
is lacking.

In this Letter, we report on PCS of single crystalline thin films of the ferropnictide superconductor 
Ba(Fe$_{0.92}$Co$_{0.08}$)$_2$As$_2$. The films are epitaxially grown on a template of epitaxial SrTiO$_3$ grown on 
La$_{0.3}$Sr$_{0.7}$Al$_{0.35}$Ti$_{0.35}$O$_9$ crystals \cite{Eom}. In the point-contact spectra, the features associated with Andreev 
reflection \cite{BTK} at the point-contact are clearly seen, but other features including a strong asymmetry in the conductance with respect to 
the voltage across the point-contact are also observed.  From a systematic analysis of the antisymmetric part of the spectra and its temperature
dependence, we show that the asymmetry might provide useful information about the rich normal state properties of these superconductors.

In order to perform PCS the sample is mounted on a home-built low temperature scanning probe microscope adapted to perform
spectroscopy in the point-contact mode. The point-contact is formed between the thin films and sharp metallic tips of silver. The microscope is 
equipped with a sophisticated coarse approach mechanism which gives us fine control on the size of the point-contact. The microscope with the 
sample is dipped in a liquid He storage dewar and the temperature is controlled by a heater and a diode thermometer mounted on the sample stage.
The measurements are done by an ac-modulation technique using a lock-in amplifier. A representative spectrum captured at 10.4 K is shown in
Fig. 1a. The raw data are clearly seen to be asymmetric about zero bias. For analysis we have extracted the symmetric and the antisymmetric 
(Fig. 1b) components of the differential resistance using the equation:
\begin{equation}
(\frac{dV}{dI})_{s,as} = \frac{dV/dI(+V) \pm dV/dI(-V)}{2}.
\end{equation}
The spectroscopic features arising in the symmetric component $(dV/dI)_s$ and their temperature evolution have been discussed in another 
publication \cite{Goutam}. Here we concentrate on the antisymmetric component. At 10.4 K the antisymmetric component of the differential 
resistance is flat with its magnitude being almost zero at lower bias for $|V| < $ 12.5 mV (Fig. 1b).  Beyond this voltage range, the magnitude 
increases smoothly with $V$ and shows a linear dependence for $|V| >$ 20 mV. The general shape of the curve changes with increasing temperature 
(shown by the color plot in Fig. 2a).  The width of the flat region in voltage first decreases; then, close to a temperature of 27 K, the flat 
region disappears and the antisymmetric component changes sign near $V = 0$ (Fig. 2c). At still higher temperatures, the flat region reappears, 
reducing in range as the temperature is increased further, but remaining until our highest measured temperature. In this higher temperature range, 
the magnitude of the overall antisymmetric component (Fig. 2d) is much smaller\cite{Surface}.

\begin{figure}[h]
\begin{center}
\includegraphics[width=8.5 cm]{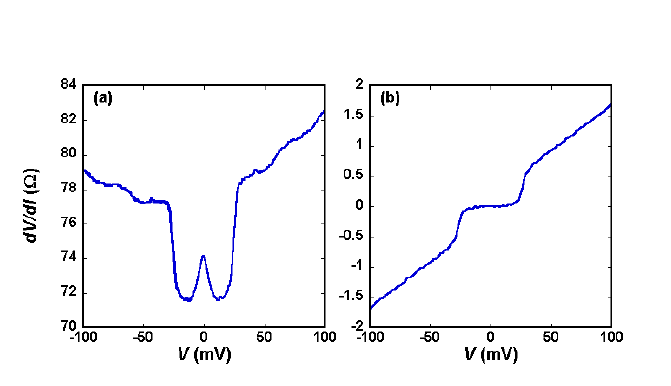}
\caption{Point-contact spectra obtained between metallic Ag and an epitaxial thin film of Ba(Fe$_{0.92}$Co$_{0.08}$)$_2$As$_2$. (a) The raw data.  
(b) The antisymmetric component of the data presented in (a) extracted following the method described in the text. 
\label{fig1} }
\end{center}     
\end{figure}

An often-quoted explanation for conductance asymmetries in point-contacts is a contribution due to thermoelectric effects arising from a 
temperature gradient existing in the point contact.  In order to understand this, it is important to briefly review the different regimes of
electron transport in metallic point-contacts \cite{thermal}. Depending on the size of the point-contact and the electronic mean free path of the 
materials forming the point-contact, electronic transport can take place in different regimes. A point-contact is in the \textit{ballistic}
limit when the effective point-contact size - contact diameter $a$ - is significantly smaller than the elastic mean free path - $l_{elastic}$. In this
regime, the electron only accelerates within the contact and dissipates energy away from the contact. If the contact size is larger than the 
inelastic mean free path {$a \gtrsim l_{inelastic}$, the electron can dissipate energy within the contact region. In this so-called \textit{thermal}
regime, the local temperature ($T_{eff}$) at the center of the point-contact can be considerably higher than that of the bath temperature $T_{bath}$.
For a point-contact between two dissimilar materials, the temperature at the center of the point-contact is given by 
\begin{equation}
T_{eff} = \sqrt{T_{bath}^2+\frac{V^2 \rho_1 \rho_2}{(L_1 \rho_2 + L_2 \rho_1)(\rho_1+\rho_2)}}
\end{equation}
where, $L_1$ and $L_2$ are the Lorenz numbers corresponding to the two materials with resistivities $\rho_1$ and $\rho_2$ respectively.  
In both cases, $T_{eff}$ goes as $V^2$ for small voltages, as should be expected, since the effective temperature  should not dependend on the 
direction of the current flowing through the contact.  Consequently, the contribution to the $I-V$ characteristic for thermoelectric effects is 
\textit{symmetric} in bias, leading to a differential voltage characteristic $dV/dI$ arising from thermoelectric effects that is 
\textit{antisymmetric} in the applied voltage.   The corresponding asymmetry in $dV/dI$ can be represented by the antisymmetric component
\begin{eqnarray}
(dV/dI)_{as} &=& (dT_{eff}/dI) (S_1(T)-S_2(T))\\ \nonumber
&\approx& (dT_{eff}/dV) (S_1(T)-S_2(T))
\end{eqnarray}
where, $S_1$ and $S_2$ are the Seebeck coefficients of the two materials respectively.  Here we assume that the differential resistance $dV/dI$ 
does not change appreciably, so that we can replace $dT/dI$ with $dT/dV$ without significant error.  This is clearly not true in the superconducting
case, but the symmetry argument is not affected. The difference in the Seebeck coefficients between the two materials can be determined from the
measured antisymmetric component of  $dV/dI$ once $dT_{eff}/dV$ is known, which can be calculated from Eqn. (2).  That is exactly what is seen 
in thermal point-contacts between two metals \cite{thermal}. Since the ferropnictide materials are known to have rich thermoelectric properties 
\cite{Sato_Thermo}, it is tempting to attribute our observation to the processes mentioned above.

However, our data are obtained from a point-contact that is either in the ballistic limit or close to the ballistic limit, but definitely not in 
the thermal limit. This is confirmed by the following observations for the symmetric component of the differential conductance $(dI/dV)_s$ 
(see ref. \cite{Goutam}): (i) the general shape of the symmetric component of the spectrum clearly shows a conductance dip at zero bias which is 
the hallmark of a ballistic point-contact on a superconductor \cite{BTK}; (ii) There is no conductance dip observed that could be related to the 
contribution of the critical current \cite{Goutam_critical}; (iii) the normal state resistance of the point-contact is high ($R_{PC} \sim$82 $\Omega$)
and does not vary noticeably with increasing temperature and (iv) The approximate size of the point-contact is estimated to be $\sim 2.5$ nm by 
putting in the normal state contact resistance in Wexler's formula\cite{Wexler} given by
\begin{equation}
R_{PC} \sim \frac{\rho l}{3\pi a^2} + \beta \frac{\rho}{a},
\end{equation}
where $a$ is the contact diameter, $\beta \sim 1$, $\rho$ is the resistivity of the material in the normal state and $l$ is the mean free path. 
The estimated size of the point-contact turns out to be smaller than the typical mean free path in usual ferropnictide crystals (lower bound to the
mean free path was found to be $\sim 3.5$ nm)\cite{MFP}.

In the ballistic limit of point-contacts, the above explanation for the conductance asymmetry is not valid as there is no dissipation of heat 
expected within the point-contact with $T_{eff} = T_{bath}$, and $T_{bath}$ is not $V$-dependent.  A more sophisticated analysis would require 
the calculation of thermoelectric voltages from the nonequilibrium distribution functions that arise in the ballistic regime \cite{Ronald}. However, this has not yet been considered for point-contacts.

\begin{figure}
\begin{center}
\includegraphics[width=8.5 cm]{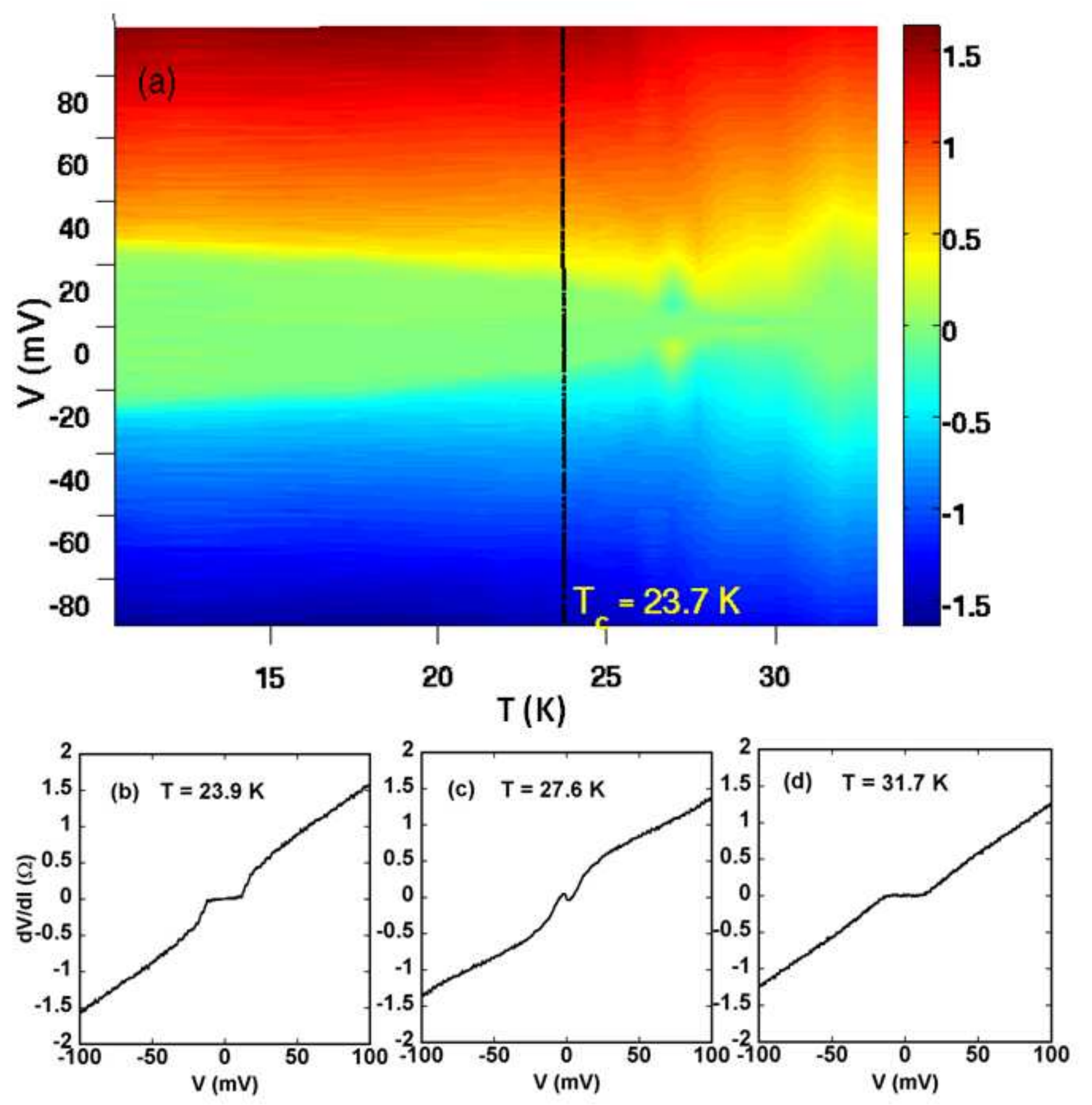}
\caption{Temperature dependence of the antisymmetric part of the point-contact spectra. (a) Color plot of the spectra between temperatures 5.8 K and
32 K. The black line indicates the spectrum at 23.7K, the mid-point of the resistive transition. (b) Spectrum at 23.9 K, just above the transition, 
(c) Spectrum at 27.6 K and (d) Spectrum at 31.7 K.
\label{fig2} }
\end{center}     
\end{figure}

Therefore, the antisymmetric component $(dV/dI)_{as}$ does not appear due to conventional point-contact heating and consequent thermoelectric effect.
However, a comparison of $(dV/dI)_{as}$ with $(dI/dV)_{s}$ at temperatures far below the superconducting transition temperature shows that the 
voltage range below which $(dV/dI)_{as}$ is flat is identical to the voltage that corresponds to the position of the conductance peak signifying 
the superconducting energy gap in $(dI/dV)_{s}$ (see ref. \cite{Goutam} and Fig. 1). Thus, the antisymmetric component is almost zero when the 
voltage across the point contact is less than the superconducting gap voltage, becomes finite just above the gap and varies linearly at higher 
voltages. The temperature evolution of  $(dI/dV)_{s}$ clearly indicates that the superconducting energy gap decreases with increasing temperature, 
but does not vanish at the critical temperature of the superconducting film ($23.7$ K in this case), surviving to temperatures greater than $30$ K. 
As discussed in \cite{Goutam}, the superconducting critical temperature of the film is smaller than the temperature where pairing of the 
quasiparticles occurs without a global phase coherence. In such a case, phase-incoherent superconducting pairs in the normal state might cause a 
pseudogap to survive well above the critical temperature of the superconductor, and this pseudogap contributes to the thermoelectric effect in 
the same way as the conventional gap. It seems that this phenomenon is prominently exhibited by the antisymmetric component of the differential 
resistance $(dV/dI)_{as}$. Survival of the flat region even above 32 K, where features due to Andreev reflection have completely disappeared 
in $(dI/dV)_{s}$ indicates that the normal-state gap survives even above this temperature. We cannot resolve this feature in $(dI/dV)_{s}$ beyond
32 K \cite{Goutam}.


In conclusion, we have observed significant conductance asymmetry in point-contact Andreev reflection spectra between metallic silver and 
epitaxial thin films of Ba(Fe$_{0.92}$Co$_{0.08}$)$_2$As$_2$. We show that the asymmetry observed in our ballistic point-contacts does not originate
from the usual generation of local thermoelectric voltage for point contacts near the thermal limit. We argue that the antisymmetric component 
provides useful spectroscopic information about the Fermi surface of the material both in the superconducting and in the normal state and associate
the unusual evolution of the antisymmetric component above $T_c$ with the existence of the superconducting gap above $T_c$ that might originate 
from strong fluctuation of the phase of the superconducting order parameter. 

This work was supported by U.S. Department of Energy, Office of Basic Energy Sciences through grant No. DE-FG02-06ER46346 at Northwestern University
and through grant no.  DE-FG02-06ER46327 at University of Wisconsin-Madison. The work at the NHMFL was supported under NSF Cooperative Agreement 
DMR-0084173, by the State of Florida, and by AFOSR under grant FA9550-06-1-0474.






\begin{thebibliography} {text}
\bibitem{Surface} Point-contact spectroscopy is known to be a surface sensitive probe. However, it is less surface sensitive than Scanning Tunneling 
Spectroscopy where only one atomic layer on the surface is probed.  In our measurements we can probe several monolayers below the surface. Therefore,
it is rational to consider our observation as a bulk property of the material.
\end{thebibliography}
\end{document}